\documentclass[preprint2]{aastex}
\newcommand{\eg}{{\it e.g.,\ }}
\newcommand{\ie}{{\it i.e.,\ }}
\newcommand{\etal}{{\it et al.\ }}
\newcommand{\degs}{^\circ}

\newcount\notenumber
\newcount\eqnumber
\newcount\fignumber
\newbox\abstr
\newbox\figca

\def\spose#1{\hbox to 0pt{#1\hss}}    
\def\ltsim{\mathrel{\spose{\lower 3pt\hbox{$\mathchar"218$}}
     \raise 2.0pt\hbox{$\mathchar"13C$}}}
\def\gtsim{\mathrel{\spose{\lower 3pt\hbox{$\mathchar"218$}}
     \raise 2.0pt\hbox{$\mathchar"13E$}}}
     
\def\gg{\hbox{$>$\hskip-4pt $>$}}
\begin{document} 
\title{\bf Dust and Stellar Populations in the Large Magellanic Cloud}

\author{Dennis Zaritsky \altaffilmark{1}}
\affil{UCO/Lick Observatories and Board of Astronomy and Astrophysics,}
\affil{Univ. of California at Santa Cruz, Santa Cruz, CA, 95064}
\altaffiltext{1}
{Present Address: Steward Observatory, Univ. of Arizona, Tucson, AZ, 85721}
\email{dennis@ucolick.org}

\begin{abstract}
We present an analysis of line-of-sight extinction measurements obtained
using data from 
the Magellanic Clouds Photometric Survey (Zaritsky, Harris, \& 
Thompson 1997, AJ, 114, 1002), which provides 4-filter photometry
for millions of stars in the Large Magellanic Cloud. 
We find that visual extinctions are typically larger by
several tenths of a magnitude
for stars with effective temperatures $ > 12000$ K,
than for
stars with effective temperatures between 5500 K and 6500 K.
Several repercussions of this population-dependent extinction are discussed. 
In particular, LMC distance measurements that utilize
old stellar populations, but use extinctions derived from OB stars, may be
biased low. 
As a specific example, we show that the LMC distance modulus
derived from field red clump stars
is revised upward relative to published measurements
by $\sim$ 0.2 mag if one uses the extinction measured for
a matched stellar population.
Conversely, measurements that utilize the youngest stars are 
subject to greater, and more variable, extinction leading preferentially
to results that may be biased high. 
Population-dependent
extinction affects the interpretation of color-magnitude diagrams
and results in an effective absorption law that is steeper than
that intrinsic to the dust for unresolved stellar systems. 
We further explore the relation between
the stellar populations and dust by comparing our extinction
map to the 100$\mu$m image of the region and identifying potential 
heating sources
of the dust. We find that although regions of high 100$\mu$m flux
are associated with young stars, young stars are not necessarily associated
with regions of high 100$\mu$m flux and that $\sim$ 50\% of the 100$\mu$m flux
is emitted beyond the immediate regions of high OB stellar density. 
We conclude that 
100$\mu$m flux should be used with caution as a star formation tracer,
particularly for studies of star formation within galaxies. 
Finally, we reproduce the observed extinction variation between
the hot and cold stellar populations
with a simple model of the distribution of the 
stars and dust where 
the scaleheight of the cooler stars
is $\gg$ than that of the dust (which is twice that of the OB stars; 
Harris, Zaritsky, \& Thompson 1997, AJ, 114, 1933).
\end{abstract}


\section{Introduction}

The distribution of dust within galaxies and its effect on 
our observations are poorly determined and have serious
implications for many areas of extragalactic astronomy. In a previous
paper (Harris, Zaritsky, \& Thompson 1997; hereafter HZT), we applied 
the classic
technique of determining line-of-sight extinction from 
the reddening-free colors of OB stars (Johnson \& Morgan 1953) to 
produce an extinction map of a $1.9\degs \times 1.5\degs$\ 
region of the Large Magellanic 
Cloud (LMC) and to investigate the relative line-of-sight 
distributions of the dust and OB stars. 
We found that the OB stars in the LMC have a scaleheight that
is about half that of the dust. The combination of this 
result and the observed spatial variations among
different stellar populations
(cf. Harris and Zaritsky 1999) 
suggests that each stellar population
will have its own distinct distribution of extinction values.
Can the relative distributions of stellar populations and
dust be mapped and what effects does this variation
have on observations of the Large Magellanic Cloud in
particular and of galaxies in general?

The correlation between dust and certain stellar populations is well-known
(\eg OB stars and dust are preferentially found in galactic spiral
arms), but generally ignored when treating the global effect of
dust on galaxy photometry due to the complexity and
ambiguity such a treatment introduces. 
One specific example where such a correlation may be important 
is found in the measurement of the distance to the Magellanic Clouds. 
Different reddenings toward
Cepheid, RR Lyrae, RGB tip stars, and red clump stars 
would result in apparent differences among
the derived distance moduli for each of these populations.
Apparent differences of
a few tenths of a magnitude in distance moduli 
are observed (eg. $m-M \sim$  18.3 for RR Lyrae and red clump stars;
Layden \etal 1996; Udalski \etal 1998; Stanek \etal 1998; Cole 1998,
Girardi \etal 1998;
and $m-M \sim$ 18.5 to 18.7 for Cepheids; 
Madore and Freedman (1998) and Feast and
Catchpole (1997)) and render the distance measurements to this key
calibration galaxy systematically uncertain to about 10\%. 
Although there are many factors that give rise to the discrepancies, 
the differences, at least
in part, may arise from applying extinctions derived from a different
stellar populations than that being used as the standard candle. 

For unresolved galaxies,
even the most detailed reddening corrections 
are fairly rudimentary (typically assuming
uniform, symmetric distributions of dust; cf. Tully \& Foqu\'e 1985).
Although 
increasingly more detailed treatments are being constructed and applied
(Huizinga 1995; Witt \& Gordon 1996; Jones, Davies, \& Trewhella 1996;
Kuchinski \etal 1998), these focus primarily on the
effect of a clumpy absorbing media and have not yet included
the additional complexity of varied distributions of dust and
stellar populations. A complete treatment 
is obviously a difficult, complicated
problem for which observed examples of dust/star 
geometries would provide invaluable constraints.
The Clouds provide an opportunity to measure the effects of a non-uniform
dust distribution on different stellar populations through
line-of-sight extinction measurements for millions of stars.

Once the distributions of stars and dust are determined, we can
begin to investigate issues regarding detailed physics of
the interstellar medium. A specific example that we can address
is the outstanding issue of how much of the far infrared
emission comes from dust heated by 
the UV flux from nearby OB stars (Young
\etal 1986; Devereux \& Young 1990) rather than
by the galactic interstellar radiation field (Lonsdale-Persson \&
Helou 1987, Walterbos \& Schwering 1987). The resolution of this question impacts the degree
to which
the far-IR flux can be used to measure, or at least trace, 
recent star formation activity.
At least two dust components are acknowledged to be
contributing to the far infrared flux:
a warm component associated with H II regions (and hence
massive stars) and a cold component, the cirrus, that is heated
by the interstellar radiation field.
Previous studies have by necessity focused 
on galaxies with marginally resolved stellar populations
(\eg M33; Devereux, Duric, \& Scowen 1997 ) or
unresolved populations (cf. Young \etal 1996; Lonsdale-Persson \& Helou 1987), 
so it has been difficult to investigate the detailed
physical connection between stars and dust emission.
The contribution from the cirrus component
is thought to depend on Hubble type (Sauvage \& Thuan 1992) decreasing to 
a contribution of $\sim$3\% in Sdm galaxies, like the LMC. 
The expectation that
little 100$\mu$m emission will be unassociated with young stars in the LMC
is simple to test because 
regions populated by young stars are easily identifiable from our photometry, 
the IRAS image has resolution comparable to the sizes of star
forming regions, and our extinction map provides
a measurement of the dust column density.

With line-of-sight extinction measurements to a
wide set of stellar populations within the LMC, we
can begin to explore the relative distribution of dust and stellar
populations, at least in this one galaxy.
This study is a necessary step to the 
accurate detailed interpretation of the rich color-magnitude diagrams
of the Clouds currently being produced.
To motivate this study in a wider context, 
we have described a few ways in which a dust map of the LMC
relative to various stellar populations may address issues in extragalactic
astronomy and ISM physics. 
The previous study of extinction in the 
LMC with the highest spatial resolution
utilized 2069 O and B main sequence stars drawn from a
dataset of over 1 million stars (HZT). To extend the range of stellar types
included in the extinction
analysis and enlarge the sample, we develop and describe
a technique that measures the extinction by fitting 
stellar models to the stellar spectral energy distribution. 
We derive estimates of both 
the reddening and effective temperature of every star in our survey 
for which 4-band photometry is available. In Section 2 we briefly
describe the data from a region of about $4\degs \times 2.7\degs$
in the LMC. In Section 3 we describe our analysis and
in Section 4 our results.

\section{Data}

These data come from the ongoing Magellanic Clouds Photometric
Survey described initially by Zaritsky, Harris, \& Thompson (1997;
hereafter ZHT) which is
a $UBVI$ imaging survey conducted with the Las Campanas Swope
(1m) telescope, the Great Circle Camera (Zaritsky, Shectman, \&
Bredthauer
1996), and a 2048$\times$2048 pixel CCD with pixel scale of 0.7
arcsec pixel$^{-1}$.
The data reduction is described in the original paper
except for one modification. 

We revise the photometric registration algorithm because the results
were unsatisfactory when the original scheme was applied to the 14
drift scans included here. 
We developed a superior alternate algorithm 
based on adjusting the photometry of the individual subscans reduced
with DAOPHOT II (Stetson 1987). 
We assume that the photometry of each subscan can be related to the
true photometry by a single zero point offset, and we assume that
on average the photometry from the survey is accurate (i.e. has a net
zero point of 0.0 mag). We calculate the relative zero point
for a particular subscan by evaluating the median of the photometric
shifts relative to all of the neighboring subscans (calculated using
stars in common within the overlap regions between adjacent subscans).
Cycling through all of the subscans, 
we identify the subscan with the largest median offset. Its photometry
is adjusted by that offset, and the algorithm begins again. We stop
iterating when the the largest subscan offset is less than
0.02 mag (a value that we judge to be below the noise of this
procedure). The principal remaining difficulty is an artifact of the
peculiar subset of data currently available. The presence of a gap in
coverage down the middle of the region (cf. Figure 1) 
renders the relative zero points
of the right and left half of the region less than ideal (and possible
issues raised by this deficiency will be discussed). The gap will
be filled as the survey progresses and the entire region described
in this paper will be surrounded by adjacent scans.

\section{Measuring Extinction}

To measure the extinction we fit three parameters,
the effective temperature, the luminosity, 
and the extinction, to the four-filter photometry of each star.
Because we only have four data values to fit per star, 
more sophisticated modeling (eg. an exploration of the extinction
curve) is unwarranted. The requirement that
a star be detected in all four bands eliminates a substantial
fraction (0.60) of the entire sample of stars, because the $U$ and $I$
band images are shallower than the $B$ and $V$ images.
Even though the procedure appears relatively straightforward
there are various subtleties that must be addressed. 

First, we must adopt a set of suitable stellar models to fit.
We use the library of models provided by Lejeunne, Cuisinier, and Buser (1997).
These models span a wide range of metallicities, effective temperature
($T_E$, and loosely referred to as the temperature throughout
this paper), and log $g$. Special care was taken by those authors 
to place the models on the correct color
scale so that the theoretical solar-abundance model yields the
observed solar $U$ through $L$ colors. This correction is particularly
important for a use such as that being described here. Of their many
libraries, we use the solar metallicity library (lcbp00.cor)
for calibration and for a test of the metallicity dependence of the method, 
the [m/H] $= -0.5$ library (lcbm05.cor) for the 
LMC analysis (the average [O/H] gas phase abundance among LMC HII regions
is $-0.53$; Pagel \etal 1978), and the [m/H] $= -1$ library (lmbm10.cor)
for a test of the metallicity dependence of the algorithm.

Second, we must adopt a conversion system between magnitudes and fluxes.
One option is to adopt the absolute flux densities of an A0V star from
an existing reference, such as Strai\u zys
(1992).
In principle, the standard star's flux
densities enable us to convert any flux into a magnitude, or 
{\it vice versa}. However, our photometric system may not be
an exact match to this particular photometric system or to that
adopted for the theoretical models. Instead, we
adopt a more self-consistent approach by defining one of the models
to represent an A0V star ($T_{E} = 9500^\circ$ and $\log(g) = 4$). 
We calculate fluxes through the our filter system  and assign
these fluxes to a 0 mag star in each filter. 
With this calibration, fluxes and magnitudes through each of the
four passbands can be evaluated for any temperature-surface gravity
model.

Third, we must adopt a function for the wavelength dependence of
extinction across the optical portion of the spectrum. Our adopted extinction
law is a standard Galactic curve (Schild 1977 and references therein), 
which for $\lambda > 2600$\AA\ is virtually
identical to the LMC extinction curve even in environments that
show significant differences blueward of 2600\AA\ (Nandy \etal 1980;
Nandy, Morgan, \& Houziaux 1984). We parameterize the
normalization of the extinction law using the extinction
in the visual band, $A_V$. Variations in $R_V$,
$A_V/E(B-V)$, are known to occur in the galaxy 
(cf. Carrasco, Strom, \& Strom 1973; Cardelli,
Clayton, \& Mathis 1989), but we cannot 
resolve such variations in the LMC with our data. 
Koornneef (1982) demonstrated
that the typical value of $R_V$ for the LMC is similar to that of the typical
line of sight in the Galaxy, 3.1 (which is the adopted value for this 
study).

The temperature and line-of-sight extinction of a particular 
star are then derived by identifying
the best fit model for that star.
The algorithm compares the observed flux in each passband
to that calculated from a model. The model is 
renormalized to minimize residuals 
and $\chi^2$ is calculated using the observational
uncertainties. We allow the flux to be renormalized to avoid assuming
a distance to the star
(which is a more critical free parameter for the local calibrating sample
to be discussed than for the LMC stars).
The residuals are calculated as fractional differences
between the observed and predicted flux 
({\it i.e.} $\Delta$mag, which weighs each filter in the determination
of $\chi^2$ more
equally than if $\Delta$flux is used). We cycle through
all available stellar models with $T > 3000$ K and  
for each model, we cycle through $0 \le A_V \le 2$
in steps of 0.01 mag. The parameters of the model with the lowest
value of $\chi^2$ are accepted as the stellar temperature
and $A_V$. 

We use local stars of known spectral type and magnitude to 
calibrate and test the procedure. Stellar data, $UBVI$ magnitudes and 
spectral types, are collected from a series of catalogs
available on-line at the Strasbourg data center 
(http://cdsarc.u-strasbg.fr/; Lanz 1986; Mermilliod 1986; Houk \& Smith-Moore 1988) for 107 HD stars. 
Using the relationship between stellar type and effective temperature
given by Strai\u zys (1982), we convert the spectral type into
an effective temperature for each
star. In Figure 2 we show the results from various tests of fitting
the stellar models to these data. In the upper left panel of the Figure, 
we have plotted the derived effective temperatures versus the known 
temperatures, which we refer to as spectroscopic temperatures,
obtained by assuming zero extinction along the line of sight
and no photometric offset between the theoretical photometric system
of the models plus our filter transmission curves
and that used to produce the published magnitudes. 

The agreement between the two temperatures
is encouraging but there are two
qualitative failures that must be corrected. 
First, the derived temperatures are
systematically lower than the spectroscopic temperatures. Second,
the deviations become large for stars with $T_E > 10000$ K. 
In the upper right panel, we plot the results allowing the algorithm
to fit extinctions. The agreement is significantly better.
There is no noticeable offset between photometric and spectroscopic
temperatures for stars with $T_E < 10000$ K and the situation has
improved somewhat for the hotter stars as well. Next we allow 
the photometric zero points of each filter to change, and we select
the best offsets by minimizing the residuals about the 1:1 line.
The results obtained with these offsets are shown in the lower left
panel. We have improved the fit at the hotter effective temperatures, although
this improvement has been obtained by construction. The photometric
offsets are modest (1\% in $U$ and $B$, 7\% in $V$ and 12\% in
$I$). These offsets may in part arise from a mismatch of the reference
model and an A0V star or from differences in the filter transmission
curves between photometric systems.
Finally in the last panel (lower right), we plot the
result of applying the algorithm (with
freedom to fit extinction and with the new photometric offsets) to data for
$UBVRI$ standards taken from the Astronomical Almanac (1981). Again, we
have converted from stellar types to $T_E$ using the conversion table from
Strai\u zys (1982). The photometric temperature agrees well
with the spectroscopic temperature over the range 3000 K to 35000 K.
The agreement between the fitted and published temperatures for this
independent sample demonstrates that the photometric offsets derived
using the first sample are generally appropriate for other data as well.
We conclude that we can recover $T_E$ to a fractional precision of
$\sim$ 10\% over this temperature range.

Although we obtain estimates of the line-of-sight extinction toward these
standard stars, there are no published measurements to compare with.
Instead, we test our extinction
measurements by artificially reddening the photometry of these 
stars and attempting
to recover the added extinction. This approach tests the degeneracy
among various stellar models plus reddening, but does not test the 
extinction law itself. 
We apply the same supplemental 
visual extinction (0.2, 0.4, and 0.8 mag) to the 107
standard stars, refit the models, and accept the best fits only if 
$\chi^2 < 3$
(as we will do with the analysis of LMC stars). In Figure 3 we plot 
a running median of the newly derived extinction minus the initial 
derived extinction (bins of 11 stars, with stars sorted by photometric
temperature) and compare the result with the input supplemental
extinction, which is plotted as a dotted line. 
For low supplemental extinction values, $A_V =$ 0.2, which corresponds
to low total extinction because the initial extinctions are generally
small, the supplemental extinction is 
recovered well across the entire range of temperatures. As the
extinction increases, we notice some serious regions of failure. 
In particular, the region of photometric $T_E$
between 7000 and 9000 K has extinction
values that are significantly underestimated for supplemental 
extinctions of 0.4 and 0.8 mag. 

To examine the failure of the method for stars of intermediate temperatures,
we plot the results for supplemental extinction of 0.8, but
sort stars by spectroscopic, rather than photometric, 
temperature (lowest panel of Figure 3). By comparing
the bottom two panels, in particular the temperature range over
which the $A_V$ determination fails, 
we can understand what has happened. Stars with large extinctions
and $T_E \sim$ 10000 K are well fit by models with 
cooler stellar atmospheres and smaller
$A_V$. This problem is a manifestation of the
classic degeneracy evident when applying the reddening-free
method in the $U-B$,$B-V$ color-color diagram. For small extinctions,
the degeneracy is avoided, but for large extinctions various
combinations
of $T_E$ and $A_V$ are possible for stars with true $T_E$'s between
7000 and 12000 K. We also see a failure mode below 5000 K for our
algorithm. We suspect
that the problem at these low temperatures 
is a combination of insensitivity to reddening
in stars that are already quite red and the effect of strong 
molecular absorption features that effectively introduce noise into
the process of matching the models to the observations. We conclude
that for the range of extinctions expected in the LMC, we should constrain
our analysis to stars with photometrically derived $T_E$'s between 5500 and
6500 K and greater than 12000 K.

We run the algorithm on the stellar catalog obtained for our 
survey region using the [m/H] $= -0.5$ models and the adopted
photometric offsets. We 
only use stars with moderate or better photometric errors ($\sigma_U \le 0.2$, 
$\sigma_B \le 0.1$, $\sigma_V \le 0.1$, and $\sigma_I \le 0.1$) and
apply magnitude cuts (exclude stars with $U$ and $B > 22$ 
and $V$ and $I >$ 21).
We obtain satisfactory fits ($\chi^2 < 3$) for 94\%
of the stars that satisfy our criteria (a total of
543,860 stars within this region). 
For our subsequent analysis we exclude stars with fitted
$T_E < 5500$ K and 6500 K $< T_E < $ 12000 K. In 
the lower allowed temperature range we have 39,613 stars and in the upper
allowed temperature range we have 106,181 stars. 

The various photometric selection criteria pose a potential 
risk of introducing unwanted biases. A limiting magnitude necessarily
implies that heavily extincted stars will be lost from the sample.
The selection based on photometric errors similarly introduces
a ``fuzzy'' magnitude limit because the errors correlate strongly with
magnitude. Except for extremely extincted stars, 
these limits are unlikely to be important for the hotter
stars because those stars are at least a magnitude brighter than the
$V$ band limit and have $B-V \sim 0$. There is greater potential for
biases in the colder sample because these stars are fainter and redder, and so 
more likely to be lost from the $U$ band data, which is our shallowest.
Upon inspection we find that although the magnitude limit ($U>22$)
is well below ($\sim 2$ mag) the red clump population, which is where
the bulk of the cold stars lie, the photometric error cut effectively
raises the limiting magnitude to $\sim$ 21 mag. To test
whether this seriously affects the recovered distribution of
extinctions we have selected a second sample of cold stars with
$0.2 < \sigma_U < 0.4$ to recover some of the excluded stars. 
The comparison of extinction values among the two samples is 
presented in Figure 4. There is no indication that high extinction ($>0.4)$
stars are being artificially dropped from the sample due to the
$\sigma_U < 0.2$ cut. We conclude that although there is no way to avoid
losing the fainter stars along
lines-of-sight with extreme extinctions ($> 1$ mag), 
selection criteria
are not significantly biasing the recovered distribution of $A_V$'s.

Using the derived line-of-sight extinction values, 
we construct an extinction map by spatially grouping the measurements. We
regrid the survey region
and begin with a map pixel that corresponds to 120 original
pixels (84 arcsec). Regions of high stellar density could be treated with
smaller
map pixels, but this choice of pixel size 
was found to be a good compromise for the 
entire survey region.
For map pixels that contain
at least three stars with extinction measurements
the median extinction value is adopted. For 
map pixels with two or fewer values, we increase the size of the
map pixel size by another 120 original pixels.
If fewer than three stars are found within this larger map pixel, we increase
the map pixel size and try again. We continue incrementing
the map pixel size until the new, larger
pixel contains at least three extinction values.
All original pixels contained within the 
enlarged map pixels are assigned the median value.
This algorithm results in a 
final map that has regions of high
and low spatial resolution, but relatively uniform signal-to-noise. 
Finally, we create a mask that 
sets any original map pixel containing zero usable stars to 
zero extinction because such pixels 
are likely to be either off the surveyed region or to be corrupted by
technical problems.

The two maps resulting
from the analysis of high and low $T_E$ stars are presented in Figure 5
(with higher extinction regions shaded darker).
It is evident that the average extinction values are larger for the hotter
stars.
We will return to this point in Section 4. We have marked, with a
vertical
line to the right of the images, the scan that is known to have 
$I$-band data taken in non-photometric conditions and we suspect that 
the feature at the far right of that scan (near the right edge) is a
result of these non-photometric data. These non-photometric data, through
the process of photometric registration discussed in \S 2, may also have 
partially corrupted the data adjacent to the top of the
non-photometric scan, which appear to have
an anomalously low $A_V$ in the upper panel of the Figure. Finally, the other 
feature we note in our preliminary discussion of the maps
is the left/right offset present in the lower panel.
We cautioned previously that the central coverage gap renders the 
relative photometric normalization of the left and right sections 
somewhat uncertain (0.05 mag)
and a slight shift in the mean extinction is seen between the two sides
for the cool stars. Other than these issues,
there is some slight structure that correlates with
subscan or scan edges. 

\section{Discussion}

The extinction maps shown in Figure 5, in particular the one derived
from the stars with $T_E > 12000$ K, show 
large coherent, filamentary structures. Most
noticeable is the broken ring of extinction in the left hand
of the survey region that surrounds the central 
stellar association of LMC-4.
Most of the high extinction regions visible
in the top panel of Figure 5 can also be seen as regions of
obscuration
in the stellar density map (Figure 1). This coincidence confirms the
features in the extinction map as real regions of enhanced extinction.
Because of the larger number of stars used, we can trace features at
finer spatial resolution than done by HZT. 
After examining the sensitivity of the algorithm to the choice
of metallicity, we proceed by comparing
the reddening map to previous reddening maps and other observations
of the same region.

\subsection{Variation with Metallicity}

Although the difference in the inferred extinction values between
high and low $T_E$ stars is 
significant, we must exclude the possibility that this difference is due
to physical reasons other than a difference in line-of-sight extinction.
One such possible reason is the metallicity difference that presumably
exists between the younger and older stellar populations.
We have selected to fit models that correspond to the current gas-phase
abundance in the LMC for all of the stars, which may not be appropriate for
the low $T_E$ stars. 
We test the technique's sensitivity to mismatched metallicities
between the stars and models by reapplying the algorithm 
using tenth-solar and solar metallicity models
on a random subset of the data. 

The resulting
differences in $T_E$ and $A_V$ as a function of $T_E$ 
are shown in Figure 6 (we plot the difference in the inferred quantity 
between a run using the standard
[m/H] = $-0.5$ models and the other [m/H] models). 
There are no systematic differences in either the effective
temperature or extinction, except possibly at
temperatures below 5000 K for $A_V$ when using the solar
metallicity models. In both plots of the derived extinction differences,
there are one or more points, within our $T_E$ range,
for which the average difference is large ($\gtsim 0.1$ mag).
For those points, we have also evaluated the median value within
those bins and plotted these values as open circles. The median, which we
use when generating our extinction maps, is consistent with zero difference.
A change of a factor
of 10 in the adopted abundance appears to have little or no net effect on the
derivation of $A_V$ and $T_E$ over large samples of stars. 
This result reflects some of the
difficulty in obtaining metallicity measurements for individual
stars from broad band colors, and assures us that 
metallicity mismatches, which are certainly present at some level, 
are not responsible for the offset seen in $A_V$ between high
and low $T_E$ stars. 

\subsection{Comparison to HZT's OB Star Extinction Measurements}

A straightforward test of the derived extinction values for the
high $T_E$ stars is a 
direct comparison with previous results. HZT compared their results
to previous studies (primarily, Hill \etal 1994, Massey \etal 1995,
Oestreicher \& Schmidt-Kaler 1996) and found that if similar
selection criteria are used the distribution of reddening values
agree. Because the HZT results are based on a subset of the
data used here, the comparison to the HZT results that we describe
below directly tests our algorithm. The agreement between the HZT 
results and previous studies address the reliability of the data.  

We match the 
1615 stars in common between the two studies and 
present the comparison in Figure 7. On average
the two measurements agree well and there is no indication of 
a bias between the two techniques, but the dispersion about the 1:1 line
is large, 0.21 mag. Some 
of this scatter is attributable to the different
photometric zero pointing technique used here and correlates with
position in the survey region. The scatter among
extinction measurements within smaller areas of the survey
is about half of the global scatter and this
reduced local scatter gives rise to the appearance of two tighter
sequences of stars above and below the 1:1 line. From this
comparison, we conclude that the uncertainty in $A_V$ for a single
star is at least
as large as 0.1 mag and possibly as large as 0.2 mag. 
The comparison demonstrates
that the reddening to any {\it individual} star is significantly uncertain
and the results from these techniques must be averaged over several
stars, as we do to produce our extinction maps, 
to mitigate these uncertainties. Nevertheless, we believe that
the newer photometric zero-pointing technique is superior and so the
current measures of $A_V$ are preferable over the HZT values. 
For the averaging process
to be physically meaningful, the region over which the
averaging is done must have homogeneous extinction properties
and the final extinction map will not resolve structures
on scales smaller than the size of the region chosen for averaging
or medianing.

\subsection{Comparison to Published Extinctions for OB Regions}

We compare our reddening values with those published
for OB regions within our survey. We have identified four Lucke-Hodge
regions ((LH 38, 54, 83, and 114; Lucke and Hodge 1970) in common 
between our survey and the studies by Hill, Madore, and Freedman (1994)
and Oey (1996). Our median values of $E(B-V)$ for stars in the four regions  
are 0.12, 0.13, 0.13, 0.14, respectively. The published values of
$E(B-V)$ are 0.15 (Oey), 
0.01 (Hill et al.), 0.16 or 0.11 (Hill et al. and Oey, respectively for 
LH 83), and 0.08 (Oey) for the four regions in the same order as above.
The published values are typically based on only 
tens of observed stars and have large uncertainties (Hill et al. quote 
uncertainties of $\sim$ 0.08). The values we attribute to Oey are
medians taken from her list of reddenings to individual stars in 
the regions. We find no systematic differences between our
values and the published ones, although the 
difference in one case (LH 54) is slightly 
larger than the uncertainty (but only by a factor of 1.5). 
This difference may be unimportant
because Hill et al. note that their data for this region
contained significant
scatter among reddenings and many negative reddening values, suggesting
that there may have been a problem with the data. Other than the
LS 54 field, the others are entirely in agreement with the published
data, although the large uncertainties preclude more detailed comparisons.

\subsection{Comparison to the OGS's Foreground Extinction Measurements}

We proceed to test whether the extinctions derived for low $T_E$ 
stars are consistent with previous studies. The only comparable study
is that by Oestreicher, Gochermann, \&
Schmidt-Kaler 1995 (hereafter, OGS) which used Galactic stars 
in the direction of the LMC to measure the foreground extinction
toward the LMC. 
We divide our low $T_E$ sample into foreground Galactic stars and LMC
stars by attributing the large vertical plume of 
stars seen in the Hess diagram at
$B-V \sim 0.6$ and $V < 18$ to Galactic
stars (cf. Figure 1 from HZT, and references therein). 
By examining the extinction values of the foreground and LMC 
populations separately and comparing them to the measured
foreground (Galactic) extinction from OGS, we
can determine whether (1) our foreground values are in agreement with
OGS, (2) whether our LMC values are consistent with (\ie
$\ge$) the foreground extinction values, and (3) whether the metallicity
difference between the Galactic and LMC populations grossly affects the
derived extinctions.

We begin by comparing our extinction values for the Galactic and
LMC low $T_E$ populations with the OGS values. 
Unfortunately, the comparison 
is not ideal because the OGS data do not sample the region as densely
as do our data. Figure 8 shows the comparison between reddenings
to individual OGS foreground stars and values inferred from an
extinction map
constructed using our foreground stars. There is no correlation
visible (although the range of values along the axes are similar). The
scatter demonstrates the large uncertainties
in the values along any single line-of-sight.

Although the scatter in individual lines-of-sight is large, over
large areas the extinction map may be reproducible and reliable.
To test this possibility, we first compare the distribution of OGS
extinction values, including all OGS stars across the entire
LMC to improve the statistics
(no detectable difference was seen when only
using the foreground stars within our survey region), to the distribution
of extinction values derived for the foreground stars in our sample
in Figure 9. Although the $A_V$ values from the
OGS data are somewhat more widely distributed
than the values we infer from our foreground population, the modes
of the two populations are similar (both $\sim$ 0.11 mag). 
The wider OGS distribution may imply either 
that the uncertainties 
are larger in the OGS data than in our data, that we 
have underestimated $A_V$ for some of the stars in our survey,
or that the region sampled
by OGS included regions of higher foreground extinctions.
The quoted uncertainties in the OGS data are mostly $<$ 0.06 mag, 
insufficient to
account for the larger spread in values. This conclusion 
is further supported by the 
rough similarity in the fraction of points with $A_V \le 0$ in both
datasets. In bulk, the statistics of the foreground extinction 
as derived from our data and OGS's are quite similar.

We compare the $A_V$ distributions for the
Galactic and LMC stars in the inset of Figure 9. 
There is little difference in the distributions except for 
the more significant tail of higher extinction values in the 
LMC stars.
This agreement
demonstrates that there are no gross differences
between the extinctions derived from the different set of stars
due to physical differences, like metallicity, and
that the LMC data are consistent with the extinction inferred from the
foreground data (from both the OGS data and our own). 

Although statistically the extinctions inferred from the OGS data
and our own are similar, we need to confirm whether we can accurately
trace the spatial distribution of dust to validate out extinction map.
To improve the statistics for the comparison between
the OGS observations and ours, we compare the
extinction map produced from the OGS data to that
obtained using all of our low $T_E$ stars
(the LMC stars should trace the foreground extinction pattern
in addition to any pattern imposed by dust internal to the LMC).
Pixel values in our low $T_E$ extinction map are tagged
according to the extinction
value given by the OGS map for the corresponding location. 
We then bin the data using the OGS extinction values and
set the bin size to include extinction values from 
at least 200 different pixels within each bin. 
The average of our extinction
values within each bin is plotted versus the OGS extinction value
in Figure 10. There is a clear correlation between
our low $T_E$ extinction measurement and the OGS data, demonstrating
that the bulk properties of the OGS map and our extinction map are
reproduced and that our
analysis is recovering the spatial behavior of foreground extinction
in this region.
There are a few discrepant values in 
bins at low values of the OGS extinction, but the pixels 
responsible are localized
in our map. Because of this spatial localization,
the discrepancy is either due to slightly erroneous photometry 
in a limited region (\eg poor aperture correction) or to
a patch of extinction internal to the 
LMC extinction that is not represented in the OGS foreground extinction map. 
We conclude that the 
strong correlation demonstrates that we recover
the spatial distribution of
extinction and that there is no dominant component
of extinction internal to the LMC that affects the low $T_E$ stars.

Despite the strong correlation, 
the relation between the OGS results and ours
is similar to that seen
in Figure 9 (our extinction values $<$ OGS extinction values). 
This offset
may be the result of a slight bias in our method when applied
to the cool stars (\eg inclusion of some intrinsically higher $T_E$
stars for which $A_V$ is underestimated or the exclusion of some
higher $A_V$ stars due to our photometric selection),
or it may be indicative of unresolved structure
in the OGS map. If areas of high extinction are small on the sky, then
a map produced from sparsely sampled data would occasionally include
large inferred 
areas of high extinction (large areas of low extinction would also
be present but would be a better representation of ``reality''). When 
such a map is compared to a map with higher resolution, a bias
in the observed sense would exist. 
We cannot determine which of these two scenarios is
responsible for the systematic discrepancy we observe, but
we conclude that to within this possible slight bias our
extinction determinations for low $T_E$ stars recover the
statistical properties and 
spatial behavior of the foreground extinction found in previous determinations
of the foreground extinction.

\subsection{Comparison to the IRAS 100$\mu$m Image}

To determine whether the medianing we perform to mitigate
the uncertainties and create the extinction map
results in a physically meaningful map, we compare our map to 
images of the far 
infrared emission. The far IR emission comes from the dust, but the 
emitted flux depends on the dust temperature and hence on the local
radiation field. Therefore, although 100$\mu$m flux is expected to 
be a tracer of dust, we do not expect it to be a complete map
of the extinction.
We have extracted the IRAS 100$\mu$m image of the area and
present it, in conjunction with the extinction map from the high $T_E$ 
stars and the density-weighted image of main sequence stars with
$V<16.5$ and $B-V < 0.5$, in Figure 11 (B3 and earlier 
main sequence stars).
We apply our mask to the IRAS image to produce an image
that exactly matches the area 
for which we have optical
data. Most of the prominent dust features are visible in both 
the extinction map and the IRAS data, which further confirms that structures
seen in the extinction map match 
true dust structures.

There are only a few instances where significant
100$\mu$m flux exists without strong extinction or where 
strong extinction
exists without significant 100$\mu$m flux. Most of the differences can be 
traced to the presence or lack of nearby young stars. For example, the
extinction map shows a long vertical filament in the upper
portion of the left half of the survey region. In extinction,
this filament does not have strong variations along the
filament. However, the 100$\mu$m emission is strongly peaked near
the lower portion of the filament, coincident with the presence
of young stars that provide the UV flux needed to heat the
dust. A more extreme example is the small, brightest knot of 100$\mu$m emission
in the upper half of the left side of the region, which has no
noticeable
extinction consequence but is coincident with a group of young stars.
Highly localized dust may be diluted in our map by the medianing we do, but
because this area includes a large number of OB stars we have not
diluted the extinction at this specific location.
These examples demonstrate that both the column density of dust
and the presence of heating sources are important in determining the 
100$\mu$m morphology, but that in general the extinction features
determined from medianing over several stars per pixel are confirmed by
the 100$\mu$m emission. 

Although strong IR emission appears to require the presence of
coincident young stars, young stars are not always indicative of 
a strong 100$\mu$m emission feature. The dominant concentration
of young stars (of both the O and B stars)
within the survey region is in the middle of the
left side of the region. However, this area is a one of the lowest
areas of 100$\mu$m flux, presumably because there is
less dust there than in other areas --- 
as suggested by the $A_V$ map. 
Perhaps the 
young stars have removed the dust along the line of sight by
blowing a bubble out of the disk plane 
in our direction (cf. MacLow and McCray 1988). 
Such a scenario would also
be in agreement with the ring-like dust morphology surrounding this region
seen both in the extinction map and the 100$\mu$m image.
Observations of the ionized gas in this region are also 
qualitatively consistent with this scenario (Hunter 1994). If the
bubble is as large in the line-of-sight direction as in the 
cross direction ($\sim$ 1 kpc), then it has blown out of the LMC disk plane.

We quantify the connection between dust, 100$\mu$m flux, and young stars 
by examining the correlation among pixel values in 
the various images. By identifying pixels in the lower panel of Figure 
11 that contain 2 or more OB stars (about 16\% of all pixels),
we find that
71\% of all OB stars are contained in these pixels, but that only 21\% of
the 100$\mu$m flux is contained within these same pixels.  Each pixel
corresponds to a projected size of 20 pc (the dust map is smoothed using
a Gaussian of 3 pixels full-width at half maximum).
Although there is a slight increase
in 100$\mu$m flux over that expected from a uniform distribution 
(21\% vs. 16\%), most of the
IR flux is not coming from the regions of highest OB star concentrations.
Pixels without a single OB star contain 54\% of the
100$\mu$m flux.
If instead we divide the region on the basis of pixels that
have higher
than median 100$\mu$m emission (12.1 MJy/sr; 50\% of the pixels),
we find that this area contains 75\% of the OB stars.
These numbers quantify the visual impression from Figure 11 that
regions with the highest 100$\mu$m flux lie near sites
of OB stars, but that the local presence of OB stars does not
imply strong 100$\mu$m emission. These arguments are complicated
by the unknown three-dimensional nature of the problemk and the
fact that OB stars can heat dust hundreds of pc away.

The relative distribution of young stars and 100$\mu$m flux within
this region illustrates why it has been difficult to resolve the
issue of the heating source of dust in other galaxies.  Although 
the 100$\mu$m flux
correlates in general with young stars (\ie more 100$\mu$m flux
implies more young stars), a substantial fraction of
the 100$\mu$m flux is diffuse (\ie coming from beyond the 
sites of OB stars) and
there are regions rich in OB stars with little or no 100$\mu$m flux. 
It is
evident from the distribution of the two components that 100$\mu$m
flux cannot be used to trace star formation in detail (even if
globally correlated), and that there
may be a large dispersion in global characteristics as well, although
we cannot address that issue with these data.
Our data directly demonstrate the scatter in local properties, but
global properties may still be correlated if the galactic
radiation field is dominated by the radiation from OB stars.
The local scatter in any relation between 
young stars and 100$\mu$m flux is evident if one considers the central region
of LMC 4 (which is currently nearly empty of dust and hence of 100$\mu$m
emission). The main arc of young stars in this region has 17
pixels that contain as many OB stars as the densest pixel corresponding to
the bright 100$\mu$m emission knot just above the arc. If there was dust
within the arc of young stars one would expect the flux to be at least
17 times greater than that coming from the knot, which is evidently
not the case. Therefore, although
large 100$\mu$m fluxes are likely to be an accurate indication of 
star formation, the converse is less likely to be true and 
100$\mu$m flux has a large dispersion as a local tracer
of recent star formation within galaxies. 
A complete treatment (UV photometry to measure ionizing flux,
optical, HI, and a variety of infrared bands) is necessary to make
further progress in resolving the full energy budget of this process, 
but is beyond the scope of the current study.

\subsection{Variation with Stellar Population}

We have demonstrated that the extinction measurements for the
low and high $T_E$ stars are in agreement with previous measurements
of extinction and emission from dust.
Therefore, we can begin to compare with some confidence
the relative distributions of dust that affect the low and high
$T_E$ stars. 
The upper and lower panels of Figure 5 demonstrate
that the line-of-sight extinctions are generally much lower for the low
$T_E$ stars. This impression is confirmed by 
the $A_V$ histograms for the two stellar populations (Figure 12). 
The trend of declining mean extinction with declining $T_E$ (and presumably
with increasing mean age) is also present
within subsets of the high $T_E$ data.
In Figure 13 we plot the $A_V$ distribution for four subsets of 
high $T_E$ stars.
There are several trends evident as $T_E$ increases: 
(1) the number of very
low $A_V$ ($< 0.1$ mag) stars decreases, (2) the peak of the
distribution narrows and stays at an almost constant $A_V$,
and (3) the tail of high $A_V$ values becomes increasingly
more significant as $T_E$ increases.
Even within the set of high $T_E$ stars,
higher values of extinction are more common for the hotter, younger
stars. To test this result,
we also plot the HZT results using the classical reddening-free
method (the Q method). Because there are fewer stars in the HZT sample, we
divide it into only two subsamples at $V=15$. As luminosity (and $T_E$) 
increase, this sample reproduces both 
the smaller relative number of low $A_V$ values and the larger relative number
of high $A_V$ values.

Even though the non-Gaussian $A_V$ tail
affects a small fraction of the total stellar
population, it can have serious effects
on certain observations and suggests 
that the dust is highly localized.
Clumpy regions of high extinction effectively
flatten the extinction curve (in the extreme case where one has only
completely optically thick clumps, some stars will be
extincted out of the sample, but the colors of the observed stars
may not change; cf. Szomoru \& Guhathakurta 1999). This flattening
of the effective extinction curve for the LMC as a whole
is countered by our observations 
that the bluer stars are more heavily extincted
than the red stars, which steepens the effective extinction curve.
The degree to which these two effects cancel depends on the detailed
distribution of dust and stars, which are still poorly
determined in detail. Lastly, because of the non-Gaussian
$A_V$ tail, general properties
of the $A_V$ distribution, such as the mode, 
may not truly reflect
the amount of light lost from the entire population 
due to the dust. 

To demonstrate the error introduced by using the mode of the
$A_V$ distribution to correct photometry, we
compare the amount of light extincted in the 
$12000 {\rm \ K}< T_E < 14000 {\rm \ K}$ sample to that extincted in the
$22000  {\rm \ K}< T_E < 45000$ K sample. One might at first believe
that fractionally the extincted light would be comparable because the 
modes of the two distributions are quite
similar (cf. Figure 13), particularly if the available data
were insufficient to significantly 
populate the high $A_V$ tail in the hotter sample.
However, using the full distribution of $A_V$'s and assuming 
that all of the stars within the temperature range have the same
luminosity (\ie no $A_V - T_E$ correlation within the bins),
we find that 29\% of the V-band light is extincted from the cooler sample
vs. 42\% from the hotter sample. As stated before, this progression
makes the effective extinction curve for the galaxy as a whole
(ignoring scattering in unresolved galaxies) steeper (\ie smaller
$R_V$). We conclude that even among high $T_E$ stars it is critical
to measure the distribution of $A_V$ to accurately
correct for extinction.

Finally, we compare the extinctions from the hot and cool
stellar populations by
comparing the extinction maps
pixel by pixel (Figure 14). Again, we bin and average (so that
each bin contains values from at least 200 pixels).
The positive correlation in Figure 14 demonstrates 
that the dust affecting the hot population also affects
the cool population, but the correlation is not one-to-one. 
To understand this correlation, we examine a model of the relative
distributions of stars and dust in which
the LMC dust is in a plane of scaleheight $\ll$ scaleheight
of the cool stars. In such a geometry, half of the cool stars are only
extincted by Galactic foreground dust. The LMC dust partially fills
the central LMC disk plane, so 
a yet-unspecified fraction of the stars behind the
LMC midplane are extincted by the Galactic foreground plus the LMC disk
and the 
remainder are seen through holes in the LMC disk and therefore only
extincted by the Galactic foreground dust. The extinction
due to the LMC disk is taken to be the extinction predicted from
the high $T_E$ stars minus the mean foreground extinction
and constrained to be $\ge$ 0.
The foreground extinction is taken to be 0.173 mag (the average from 
the extinction map constructed from the OGS data for the relevant region of the
LMC), and we plot in Figure 14 the results of adopting dust ``filling
factors'' for the LMC dust of 0.0, 0.5, or 1.0. A filling factor
of 1 corresponds to the case where 
the disk extinction is exactly that given by
the high $T_E$ extinction map, while a filling factor of 0.5 corresponds to 
the case where half of the lines-of-sight are clear and half are extincted
as described by the high $T_E$ extinction map.
This simple model with a filling factor of 1.0 
successfully reproduces the behavior in the Figure.
The extinction predicted by the model toward the left of the
Figure is lower than observed, but only by 0.02 mag. 
If one adds 0.02 mag to the foreground model (to resolve this disagreement), 
then the models with a filling factor $\sim$
0.5 work equally well. In either case,
we conclude that a model where the low $T_E$ stars have a
significantly 
larger scaleheight than both the OB stars and the dust 
and where the dust structure recovered from the high $T_E$
stars is taken to represent that of the midplane sheet 
works well in reproducing both the low and high $T_E$
extinction behavior.

\subsubsection{Some Implications for the Distance Scale}

One area where accurate extinction corrections are 
critical is distance scale work.
The population-dependent extinction
observed in the LMC is also presumably present in 
other galaxies and can influence observations of those
galaxies. As such, it becomes important not only to measure
the extinction {\it local} to the source of interest
(\eg Cepheid or supernovae), but also to measure the
distribution of extinctions for a {\it correspondingly 
similar} stellar population. Both of these goals obviously
become increasingly difficult the more distant a galaxy is,
but they can even be important for studies of the LMC.
Inappropriate reddenings adopted using mismatched stellar populations
can affect the derived distance to the LMC, which is often
a key calibrating galaxy. Reddening estimates are generally
determined from OB stars (cf. HZT) which are bright and for which
unambiguous reddening measurements can be obtained using the straightforward
application of the classical reddening free color technique (the
Q method). However,
the results in this paper illuminate 
several key difficulties with this approach.
First, the reddenings determined in this way should not be applied
to Pop II standard candles (cf. red clump stars and red giant tip stars).
Second, even among the high $T_E$ stars, there can be differences
in the distribution of extinction values.
Third, because the distribution can be highly non-Gaussian, significant
errors can be introduced by using certain statistical properties
(\eg mode) rather than the entire $A_V$ distribution to correct
the photometry.

We do not intend here to provide a complete review of how 
variable extinction will affect each distance method. However, a few
examples will demonstrate that the magnitude of the effect is 
comparable to the degree of disagreement between various published results.
First we consider the use of the LMC to zero-point the Cepheid
P-L relation. We have already demonstrated in Figure 13 that 
even among stars with $T_E > 12000$ K (the young stellar population
and therefore appropriate for Cepheids) there are differences in the 
distribution of extinction values. Any determination of the extinction
will be dominated by the coolest stars included because there will
be many more of them. For example, the mean visual extinction obtained
using all of the stars with $T_E > 12000$ K is 0.42 and the
mean using only those stars with $T_E > 22200$ K is 0.60. Even in the I band,
where the effects of extinction are more moderate, this difference
corresponds to 0.11 mag. Assuming that we have chosen the appropriate
subpopulation to match to the Cepheids, the non-Gaussian tail creates
significant differences between the mean and median $A_V$. 
For example, for the $T_E > 22000$ K sample the mean
and median $A_V$ are 0.57 and 0.44, respectively. Which should one 
apply?
Because of the unknown small scale structure of the dust, 
a photometric correction will be statistical and the properties of the $A_V$
distribution are critical. 
As discussed in detail by Madore and Freedman (1998), the issue
of extinction in relation to the distance scale 
has been extensively investigated and the most promising
techniques are those that sidestep the issue by either using reddening-free
quantities or by moving as far into the infrared as possible
to minimize the effect of 
extinction. Nevertheless, investigators are often left with no option
but to estimate the extinction from ``local'' stars. We have shown
here that this approach has possible systematic problems.

Our results suggest another way to mitigate the influence of dust.
Because the extinction is lower for the colder, older stars, using Pop
II standard candles is analogous to going to the infrared for Pop
I standard candles.
In a previous study (Stanek, Zaritsky, \& Harris 1998), 
we used our survey data to investigate the small distance modulus,
18.08, obtained by a previous study of field red clump 
stars (Udalski \etal 1998). 
Using the map of HZT we were able to select low extinction regions
and then correct the photometry for the extinction that was present
($A_I = 0.31$,  $E(B-V) = 0.17$). That analysis confirmed the
small distance modulus determined by Udalski \etal. Since that study,
the discussion in the literature has focussed on the magnitude
of the age and metallicity corrections to the red clump magnitude. 
Corrections derived from stellar models (Cole 1998; 
Girardi \etal 1998) revised the red clump distance modulus to 18.3, which is
still below the ``standard'' value of 18.5 but within the
uncertainties. The difficulty with theoretical corrections is that they
are based on models of a complex phase of stellar evolution. 
Empirical studies of the red clump
conclude that the age and metallicities are more moderate (Udalski
1998a, 1998b) and that the distance modulus is still $\sim$ 18.1.
The difficulty with empirical corrections is that they depend on
the complicated interplay of systematic uncertainties in the various
distance estimates used to calibrate the age and metallicity
dependence of the red clump luminosity.
Our study does not address the age and metallicity corrections,
but rather returns to the issue of the extinction correction.

The HZT extinction map Stanek \etal used is based
on observations of OB stars and here we find that such a map is not an 
accurate representation of the reddening to red clump stars. 
Our low $T_E$ extinction map is ideal in correcting for the extinction
because red clump stars are within the $T_E$ range included in 
the construction of this map.
Our new data suggest that $A_V$ for the field red clump stars in 
regions of low extinction (as selected by Stanek \etal) is $\sim$ 0.1
and so $A_I \sim 0.06$, which increases the derived distance modulus
by 0.25 mag to 18.32 for the data presented by Stanek \etal. If
age and metallicity 
corrections of the magnitude suggested by Cole (1998) or Girardi \etal
(1998) are then applied ($\sim 0.2$ mag), 
the red clump distance modulus can be made as
large as 18.5, in excellent agreement with the ``standard''
value. Whether the latter corrections are necessary is not addressed by 
our data and should not be inferred from the agreement with
the standard distance value.

We do not want to leave the reader with the impression that 
the correction for population-dependent extinction removes all of 
the discrepancies among LMC distance determinations.
There remain several arguments for an LMC distance modulus that is 
smaller or larger than the standard.
Some analyses of RR Lyrae observations suggest a small LMC 
distance modulus (Walker 1992a; Layden \etal 1996),
and an erroneous reddening correction is not responsible for artificially
decreasing the apparent distance
because the reddenings for cluster RR Lyrae are 
derived from color-magnitude diagrams. The extinctions to these
clusters are an accurate measure of the extinction toward the RR Lyrae 
(cf. Walker 1992b for an example).
Only two of the seven clusters used by Walker 1992a
have $E(B-V) > 0.1$, in agreement
with our generally low extinctions to older populations. 
However not all RR Lyrae measurements distances are small, 
RR Lyrae distances using Hipparcos-based calibrations 
have given much larger distance moduli (18.65, Reid 1997; 18.63
Gratton \etal 1997). Some other methods, for example the
analysis the SN 1987A ring, which again
is not affected by the problems discussed here, can
result in a smaller distance modulus
(cf. $\ltsim$ 18.37 (Gould and Uza 1998)), but can also result
in larger values (cf. 18.5, Panagia \etal 1991; 18.43, Sonneborn \etal
1997). Specifically for the red clump stars, Udalski (1998b) presents
data from LMC clusters, for which the extinctions are small, and obtains
a low value of the distance modulus that is in agreement with 
field red clump distance modulus with the original extinction
correction. The disagreement between the field and cluster red
clump distances that results from our new extinction correction is not
easily resolved. The clusters have low extinctions, so even if the
adopted extinction is incorrect, 
the effect on artificially depressing the distance modulus is minor. Presumably
the cluster stars 
have similar ages and metallicities to the field stars, so the
resolution will not come from differential age and metallicity corrections.

These unexplained discrepancies still render the distance measurement
to the LMC uncertain. However, 
we conclude that population-dependent extinction is yet another
issue that must be addressed as one measures distances and 
that extinction differences between populations
can result in errors
comparable to other, better appreciated, sources of systematic error.

\subsubsection{Some Implications for Studies of Stellar Populations}

The results presented here also have implications for the observation
of stellar populations in both resolved and
unresolved galaxies. The study of resolved populations has
been advanced in the last few years by deep HST color-magnitude
diagrams with many thousands of stars, and ongoing ground-based
surveys that
will provide colors and magnitudes of millions of stars. 
Aside from identifying rare stellar populations that can provide
key clues to the star formation history of the galaxy, the large
samples allow precise determinations of the apparent positions
and dispersions of populations, like the red giant branch, which in
turn can constrain the range of ages and metallicities present in the
system. The 
dominant source of error in such analyses will no longer be statistical
but systematic. The extinction correction is a principal potential source
of such systematic uncertainties. We were in fact driven to this
analysis by our inability to fit the main sequence and red giant
branch simultaneously when applying a single distribution of 
extinctions to a portion of our
LMC data. Any detailed, precision analysis of stellar systems that
includes relatively recent star formation must account for 
population-dependent extinction.

The situation is even more complex in unresolved galaxies.
As previous studies have stressed (cf. Witt \& Gordon 1996), the
interpretation of the measured surface brightnesses 
is complicated by light scattered back into the
line-of-sight and multiple scatterings. 
Neither of these issues are addressed by the current data
on the resolved LMC
because any scattered light is removed as local background to each star. 
Ignoring scattering, we can estimate the amount of light lost due
to extinction from both the hot and cool stellar populations in the
LMC to determine how an unresolved galaxy with such stellar populations 
might be perceived. In the $V$ band, extinction has removed 28\%
of the light from the hot stars but only 13\% of the light from
the cool stars. In the $B$ band, extinction removes 33\% 
of the light from the hot stars but only 16\% of the light from
the cool stars. Correcting the apparent luminosity of the galaxy
using the extinction appropriate for OB stars (or alternatively from
emission line extinction measurements (H$\alpha$/H$\beta$), which
are likely to be similar to the OB values), will overestimate the
galaxy luminosity by $\sim 15$\% and introduce a color correction
that is about 2 times larger than that appropriate for the bulk
of the stellar population. Although investigators have been
aware of such issues, the difficulty in dealing with the large
unknown number of parameters has led to very simple treatments. Our
discussion here provides an indication of the degree of uncertainty
that such simplifying assumptions may impose.

\section{Conclusions}

We have fit colors derived from stellar models to 4-filter photometry
of LMC stars to derive effective temperatures and visual extinctions.
We find that there are two populations of stars for which we 
can reliably recover the visual extinction, stars with 5500 K $< T_E < $
6500 K and stars with $T_E > 12000$ K. Various tests demonstrate
that we are not strongly sensitive to metallicity and that we recover
extinctions in agreement with previous studies to within a
dispersion of $\sim$ 0.1 to 0.2 mag
for individual stars. More precise extinction values are derived by
combining the results from many stars and producing extinction
maps of the region. Our principal result is that there is a significant
difference in the extinction values derived for the two samples of 
stars, with the hotter stars being more highly extincted on average.

The difference in extinction properties as a function of stellar
population is not entirely unexpected, but we present direct evidence
and quantify the effect for the LMC. We test for
biases by comparing to previously published extinction measurements
and to 100$\mu$m images of the
region. We find that in general the 100$\mu$m emission features 
trace the extinction morphology, but that there are a few exceptions.
These few exceptions can be traced to the presence or absence of
young stars that heat the dust. 
We identify several locations where young stars have
heated the dust so that it emits strongly at 100$\mu$m and locations
where there are young stars, but apparently no dust because
there is no 100$\mu$m emission. We conclude that significant 100$\mu$m
flux correlates with young stars, but that young stars do not necessarily
imply significant 100$\mu$m flux and
that 100$\mu$m flux should only 
be used as a star formation tracer with caution. This is in agreement
with the results of Walterbos and Schwering (1987) for M 31.

We presented a simple model for the distribution of stars and dust
along the line of sight. Previous results (HZT) demonstrate that
the scaleheight of the dust is about twice that of the OB stars.
Because the
low $T_E$ stars have different extinction properties, we hypothesize
that the scaleheight of this populations is $\gg$ than that of the
dust. Such a model implies that half of the low $T_E$ stars are 
completely in front of the LMC dusty disk and that half are extincted
by the extinction at that location in the disk (given by the high
$T_E$ extinction map). 
We then compare the expectation from this model with the
spatial correlation between low and high $T_E$ extinction measurements.
We find that this model fits the data remarkably well and
provides a simple explanation for the different extinction properties
of the two samples.

We explore some ramifications of the population-dependent extinction.
In particular, we discuss its effect on distance
measurements of the LMC. The expected behavior is in the correct sense
to help resolve the discrepancy between distance moduli found using various
stellar populations (cf. RR Lyrae, red clump stars, and Cepheids).
We present how the distance measured using field red clump stars may have
been biased downward by using reddening measurements obtained from
OB stars, rather than from red clump stars, in the same region. 
Taking the mean extinction obtained from the cooler
stars, which are directly comparable to the red clump stars in temperature,
the distance modulus inferred is 0.2 mag larger
than previously derived, in better
agreement with values obtained from observations of Cepheids.
However, the Cepheid distance measurements may also be affected 
by population-dependent extinction variations because
young stars are subject to greater and more variable extinction,
making determination of the $A_V$ distribution, rather than a mean
or median, key in properly accounting for the effects of extinction.
We discuss how observations of galaxies in general may be affected
by the population-dependent reddening. Treating this section of the LMC
as a complete independent galaxy, we derive that nearly twice as much
of the light from the hotter stars is removed by extinction than of
the
cooler stars. This effect results in an effective extinction curve steeper 
than the true curve. 

We conclude by stressing that dust affects different stellar 
populations differently. Simple models of obscuring planes or
well-mixed stars and dust do not accurately represent the physical
situation. Whether this gravely affects our interpretation of 
observations is still not fully understood. We illustrate that 
in certain cases ignoring the results presented here could have
serious consequences. We continue to reduce the data obtained for
the LMC and SMC as part of our survey and soon should be able to examine
the distribution of dust on even larger scales. Then we can address
whether the lack of dust in the center of the starforming
region in LMC-4 is common for bubbles and superbubbles, 
how the dust is distributed though the
disk of the LMC, how it is distributed in the bar, and whether
similar distributions are observed in the SMC.  These data provide
a unique opportunity to explore the relative distributions of stars
and dust within galaxies and are a necessary first step in properly
interpreting the highly populated color magnitude diagrams that are
becoming available for the Clouds.

\acknowledgements

We thank Rob Kennicutt 
and Ann Zabludoff for detailed comments and suggestions.
DZ acknowledges financial support from an
NSF grant (AST-9619576), a NASA 
LTSA grant (NAG-5-3501), a David and Lucile Packard Foundation
Fellowship, and an Alfred P. Sloan Fellowship.

\clearpage

\onecolumn

\begin{figure}
\vskip 5in
\includegraphics{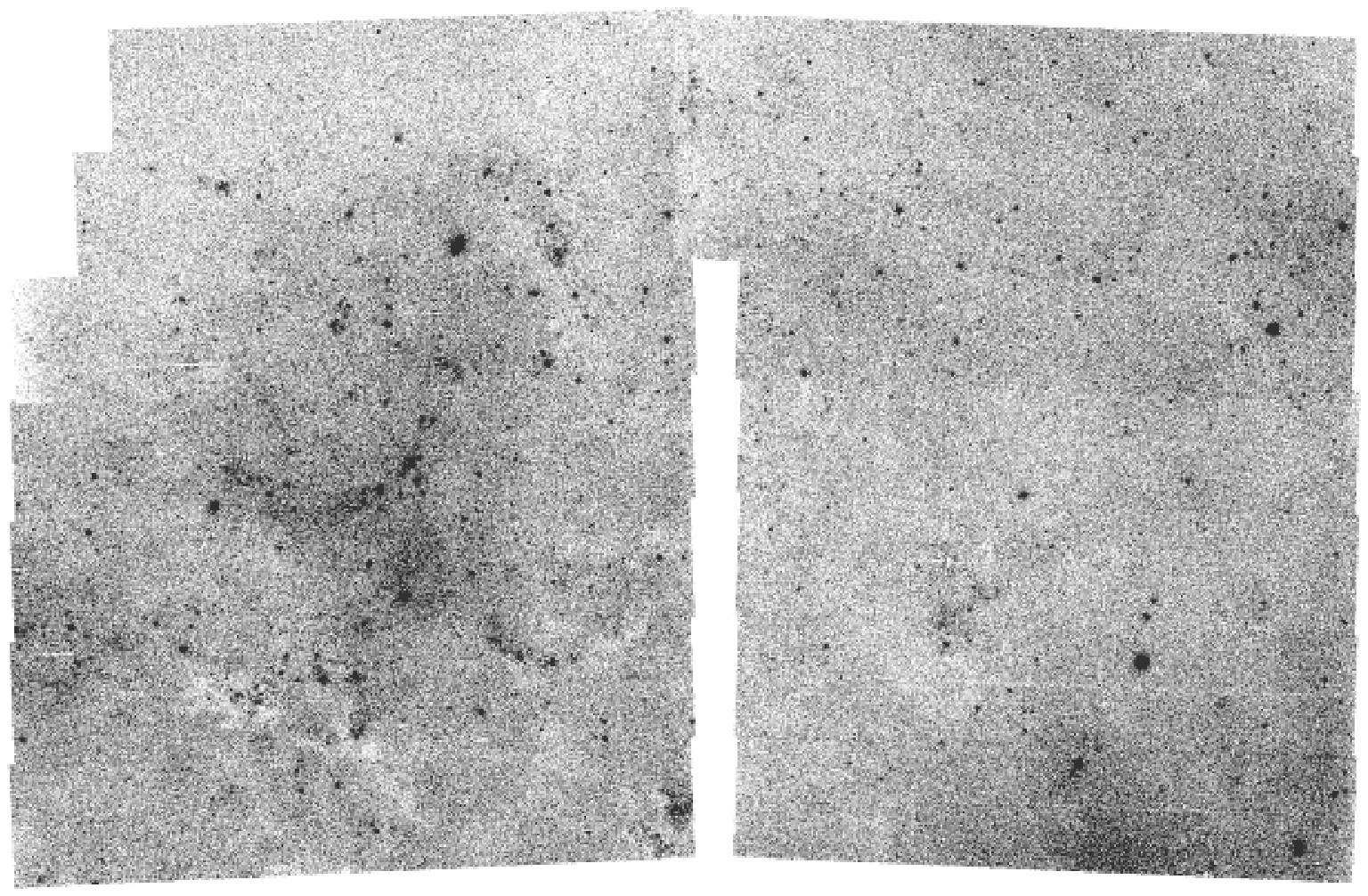}
\caption{The stellar density plot (for stars with $V < 21$) for the region
in the LMC with reduced photometry from the Magellanic Cloud
Photometric Survey. The 
central coordinates are roughly $\alpha = 5^h20^m$ and 
$\delta = -66^\circ48^\prime$,
the image is $\sim 4^\circ$ wide by $2.7^\circ$ tall, 
with North at the top and East
to the left. Each ``pixel'' in this plot corresponds to 21$^{\prime\prime}$.
\label{fig1}}
\end{figure}

\clearpage

\begin{figure}
\vskip 5in
\includegraphics{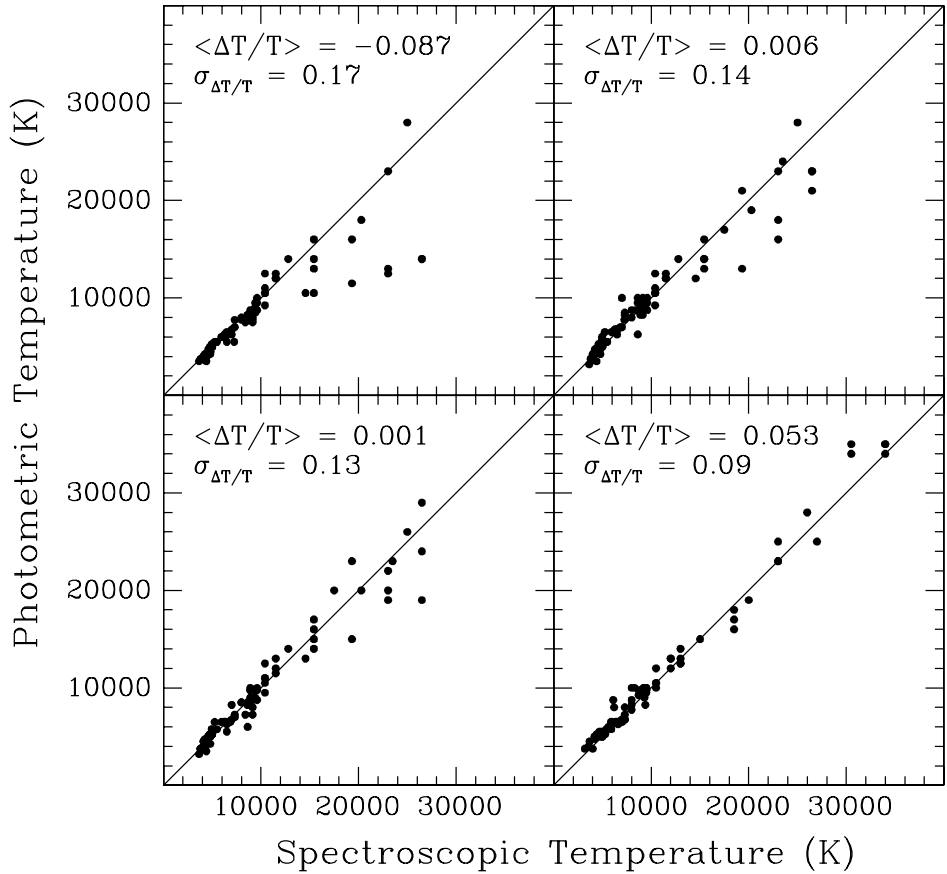}
\caption{Comparison of photometrically 
derived effective temperature with spectroscopically
derived effective temperature. Each panel depicts the results for
standard stars from
the application of an algorithm that matches stellar models to data.
The results in the upper left panel were obtained
with neither a correction for extinction 
or a photometric zero point shift between the standard stars and
the models. The results in the upper right panel 
were obtained from models that include extinction as a fitting parameter, 
but do not include a zero point photometric shift. The results in the
lower left panel are from models that include extinction 
and photometric shifts. The results in the 
lower right are from the application of models that include
extinction and the photometric shifts to an independent set of
standard stars. In the upper left of each panel 
we provide the fractional shift between 
the two temperatures and scatter from the 1:1 line.}
\end{figure}
\clearpage

\begin{figure}
\vskip 5in
\includegraphics{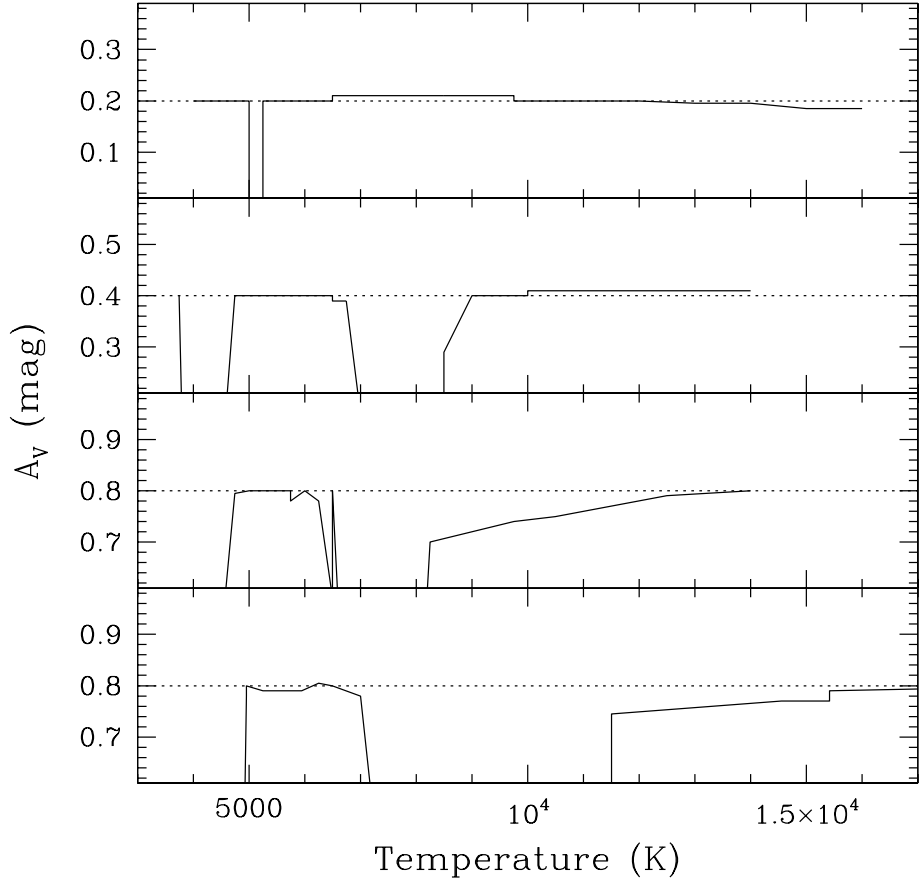}
\caption{Recovering a supplemental, artificial line-of-sight extinction. The 
top three panels
contain the recovered extinction ($A_V$) as a function of 
recovered effective temperature (a running median in $T_E$ for 11
stars) drawn as a solid line. 
To the existing data,
we added a visual extinction of 0.2 mag to
generate the upper panel, 0.4 for the second panel, 0.8 for
the third panel. Each panel
has a dotted line that corresponds to the level of supplemental, artificial 
extinction. The lowest panel shows the extinction recovered from the 
simulation with a supplemental extinction of 0.8 (same as the third 
panel) but the data are plotted versus spectroscopic $T_E$. }
\end{figure}
\clearpage

\begin{figure}
\vskip 5in
\includegraphics{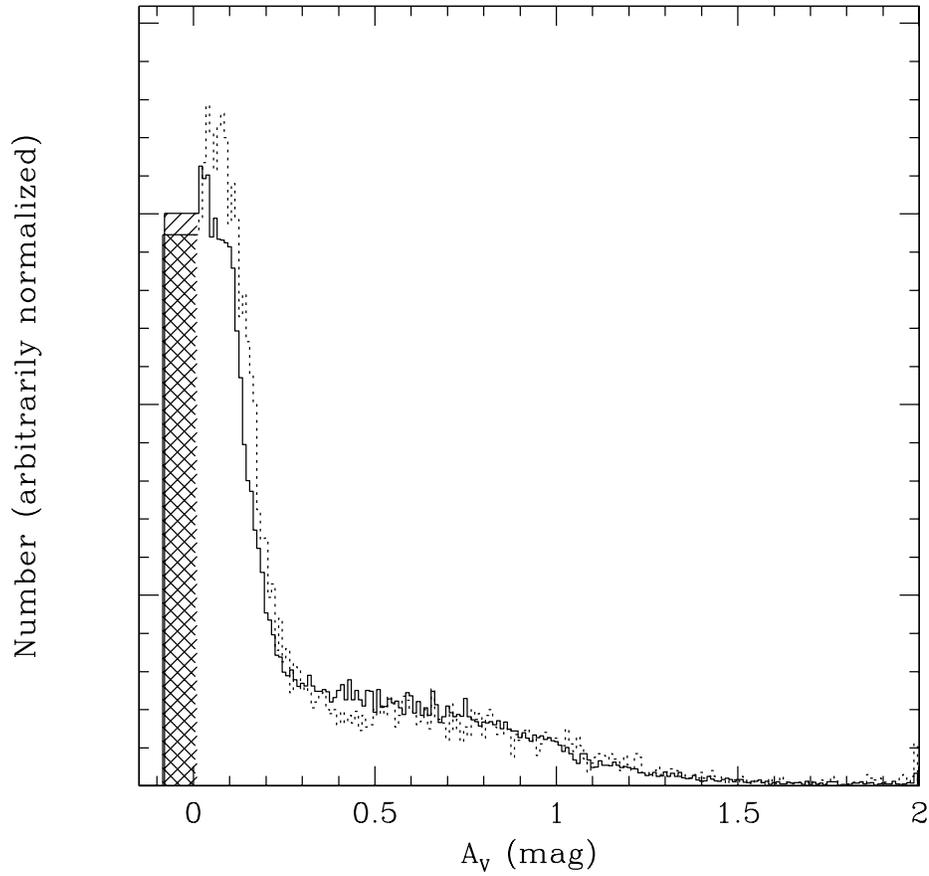}
\caption{
The distribution of visual extinctions for the original sample
of low $T_E$ stars (solid line) and for the corresponding sample with
larger photometric errors ($0.2 < \sigma_U < 0.4$; dotted line).
The hashed areas represent the stars for which the derived $A_V
=0$. The two histograms  have been arbitrarily normalized to match
at $A_V = 1$. The original sample is 3.4 times larger than the high
$\sigma_U$ sample.}
\end{figure}
\clearpage

\begin{figure}
\vskip 5in
\includegraphics{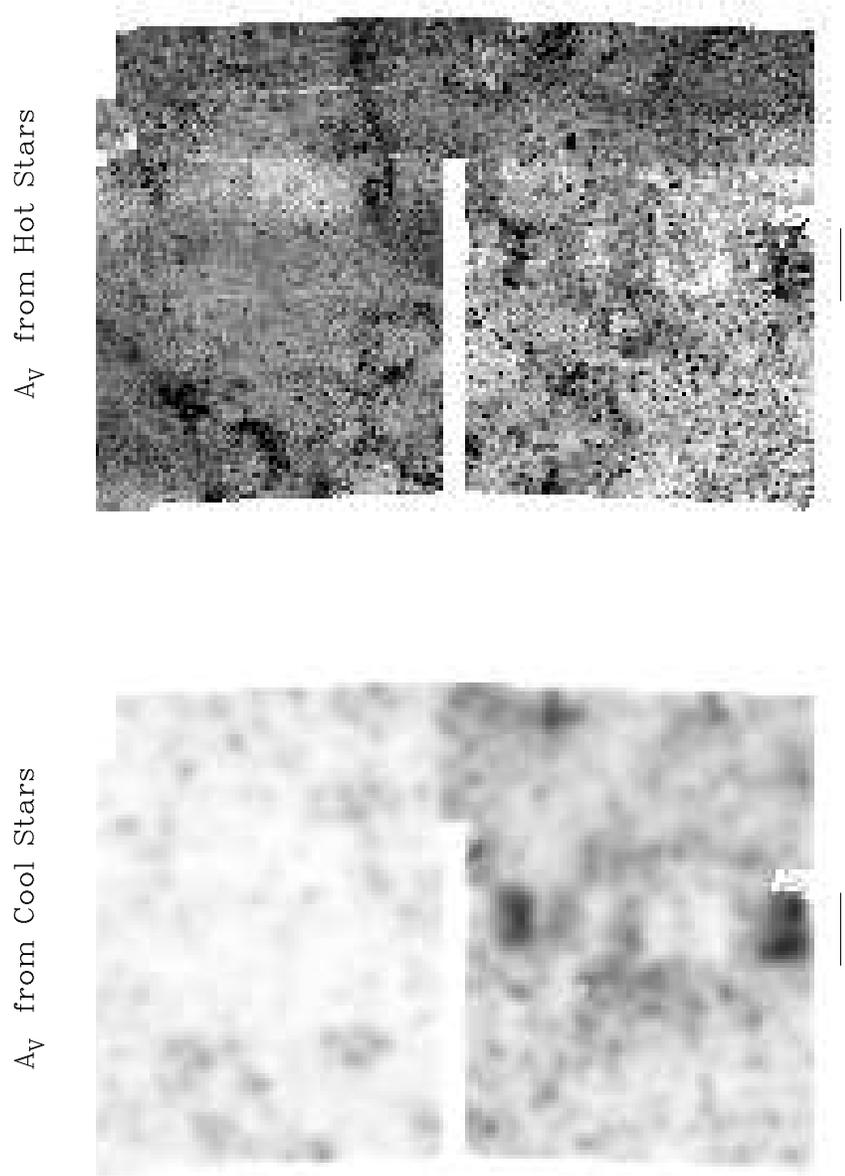}
\caption{
Extinction maps from stars with 12000 K $< T_E <$ 45000 K (top) 
and stars with 5500 K $< T_E <$ 6500 K (bottom) for the region
showed in Figure 1. The lower image
has been smoothed with a Gaussian of 1.5 map pixel dispersion to
clarify any structures. The brightness of 
the images are identically scaled (from $A_V
=0$ (white) to $A_V = 0.75$ (black)). 
The small vertical bar to the right of the images
marks the position of a scan known to have non-photometric $I$-band
data. }
\end{figure}
\clearpage

\begin{figure}
\vskip 5in
\includegraphics{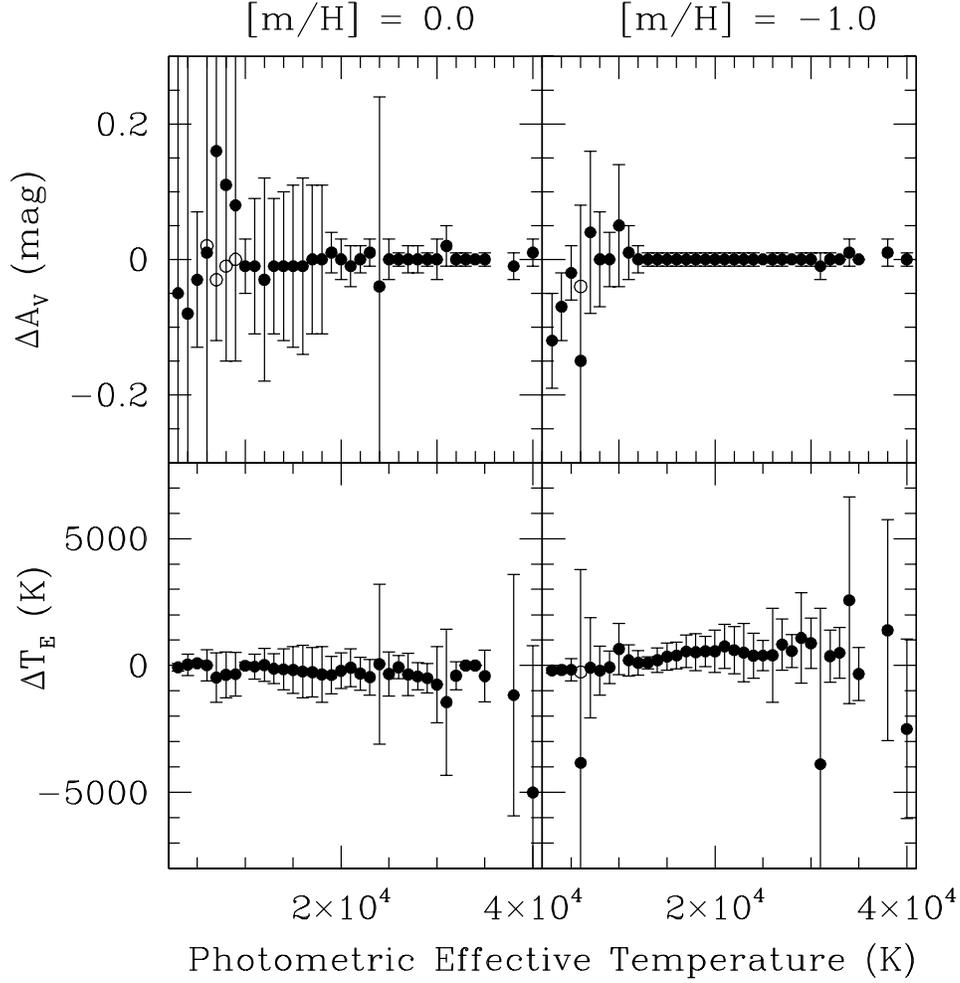}
\caption{
The dependence between metallicity and the derived extinction and
effective temperature. The mean and dispersion 
of the difference between recovered quantities for various models are
plotted. The comparison of results using standard 
models ([m/H] = $-$0.5) to solar metallicity 
models is shown in the left panels. 
The comparison to tenth-solar metallicity models in the right panels.
The open circles at 
$T_E \sim 6000$ K represent the median values in those bins (see text
for discussion). Errorbars are the dispersion of the mean.}
\end{figure}
\clearpage

\begin{figure}
\vskip 5in
\includegraphics{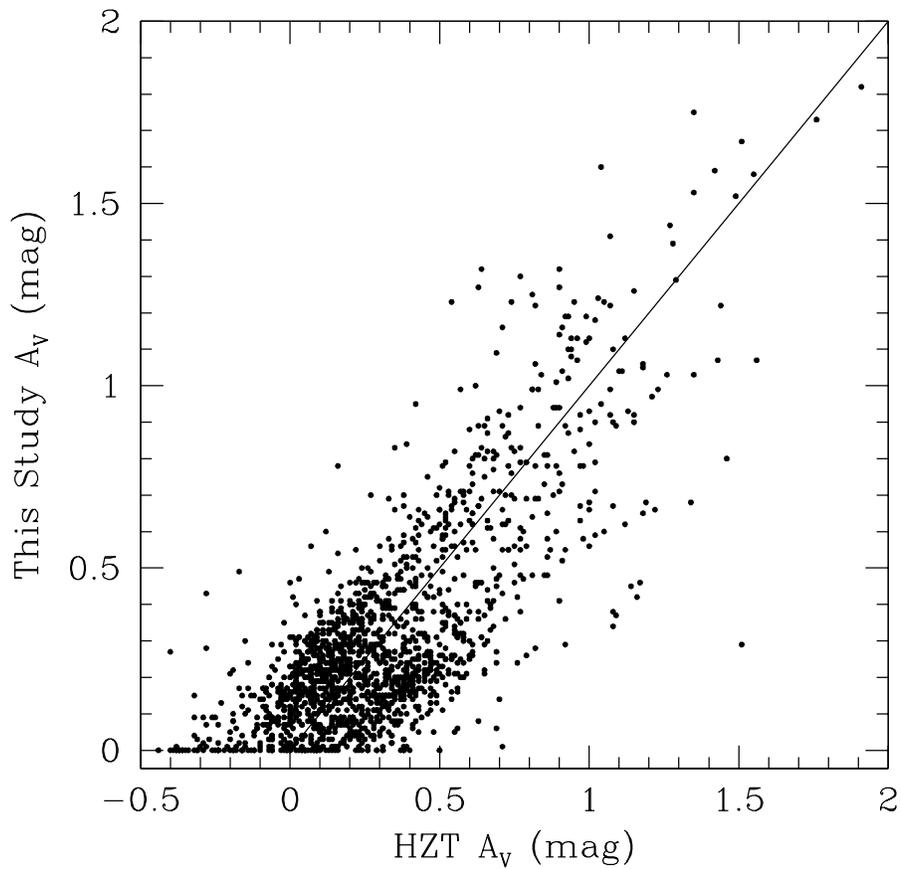}
\caption{
A comparison of extinction values derived for stars in common with
the HZT study. We compare our values to the values they derived using
the classic UBV reddening-free method. The line indicates the
one-to-one correspondence and is not a fit to the data.}
\end{figure}
\clearpage

\begin{figure}
\vskip 5in
\includegraphics{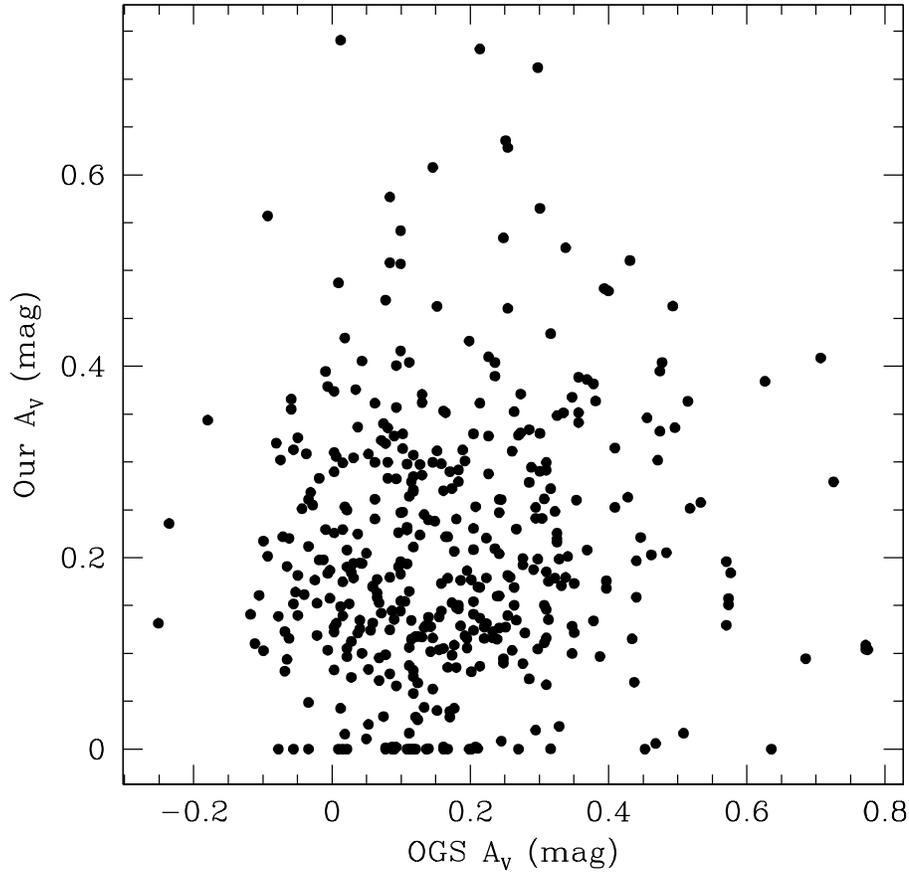}
\caption{
A comparison of the reddening measured by OSG to Galactic stars and
the foreground reddenings implied by our map at the positions of the
stars. Although the mean and
dispersion of the two populations (i.e. the global statistical
quantities) agree well, there is no discernible coherence in the 
point-by-point comparison. }
\end{figure}
\clearpage

\begin{figure}
\vskip 5in
\includegraphics{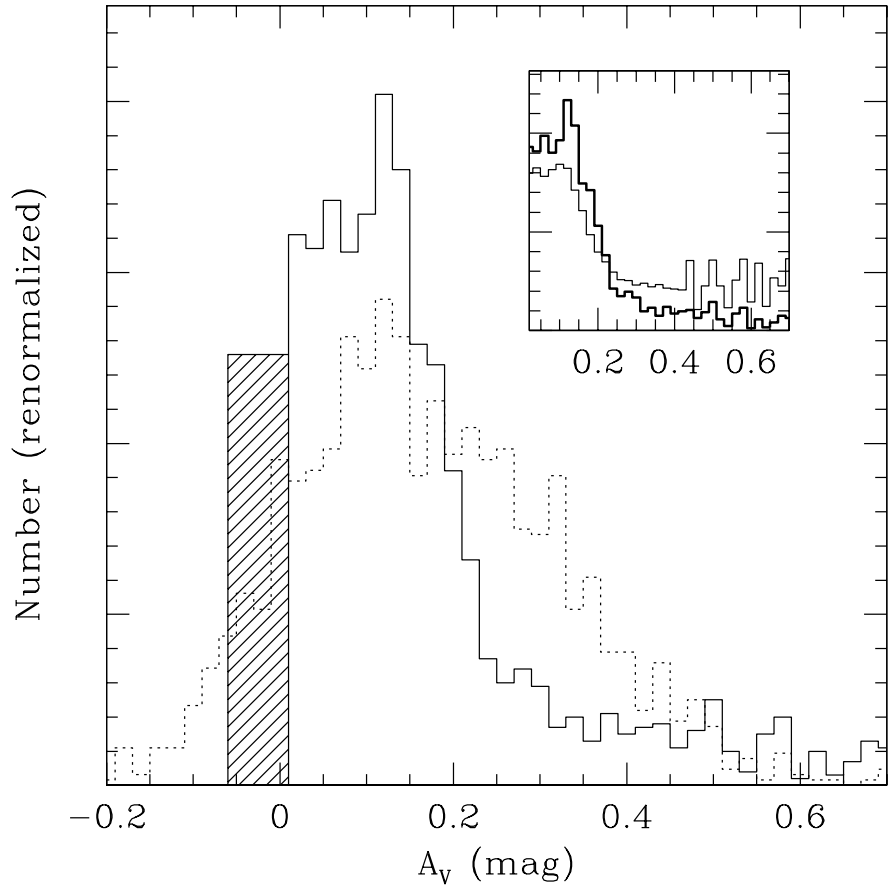}
\caption{
A comparison of the distribution of 
extinction values derived for the subset of our cool stellar population,
5500 K$ < T_E <$ 6000 K (solid line) that is interpreted to be
predominantly Galactic foreground stars 
and for Galactic foreground stars (dotted line) from OGS (1995) over their
entire LMC region.
The hashed box represents the number of stars with $A_V = 0$ in
our analysis ($A_V < 0$ is not allowed in our models). The histograms 
are arbitrarily normalized.
In the inset we present the $A_V$ distribution for both our
Galactic foreground stars (thick line) and LMC population (thin line).}
\end{figure}
\clearpage

\begin{figure}
\vskip 5in
\includegraphics{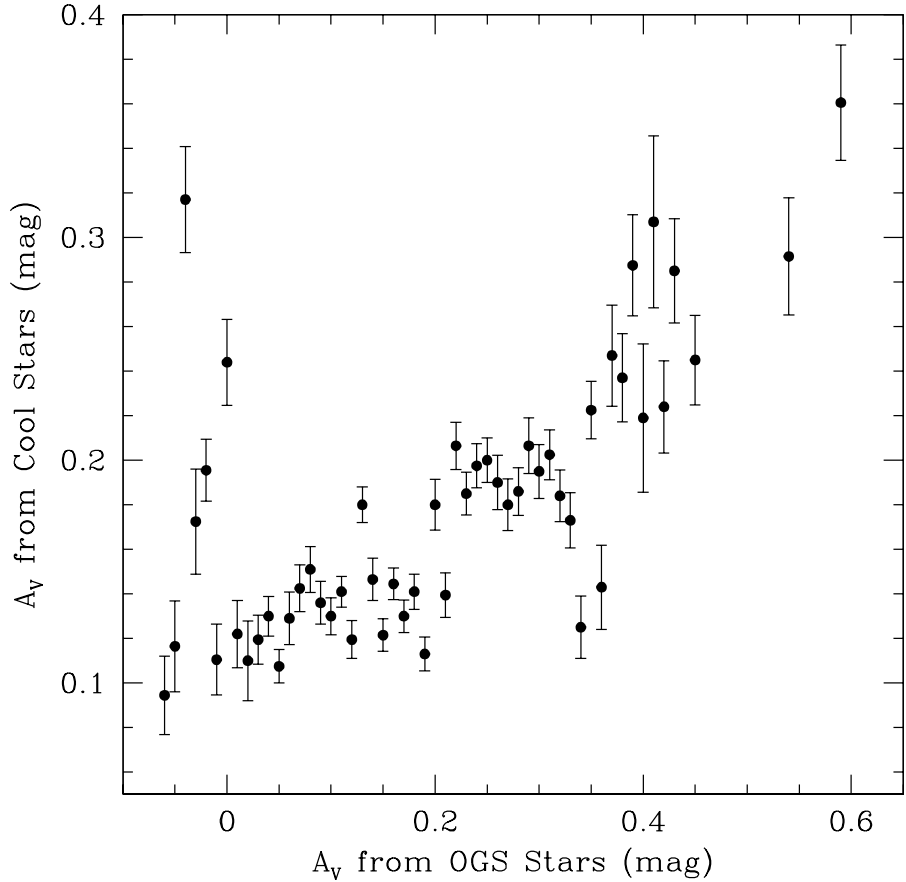}
\caption{
A comparison of the foreground reddenings determined from a map
constructed from the OSG data and our extinction map
constructed from all of the cool stars (LMC and Galactic).
Pixel values are binned according to the OGS extinction value
and averages of the pixels from our map are plotted (errorbars
are the dispersions of the mean).
The correlation is evident, indicating that structures seen in 
the Galactic foreground map are also evident in the LMC data, despite
that lack of an apparent correlation in the extinctions of individual stars
(Figure 7).}
\end{figure}
\clearpage

\begin{figure}
\vskip 5in
\includegraphics{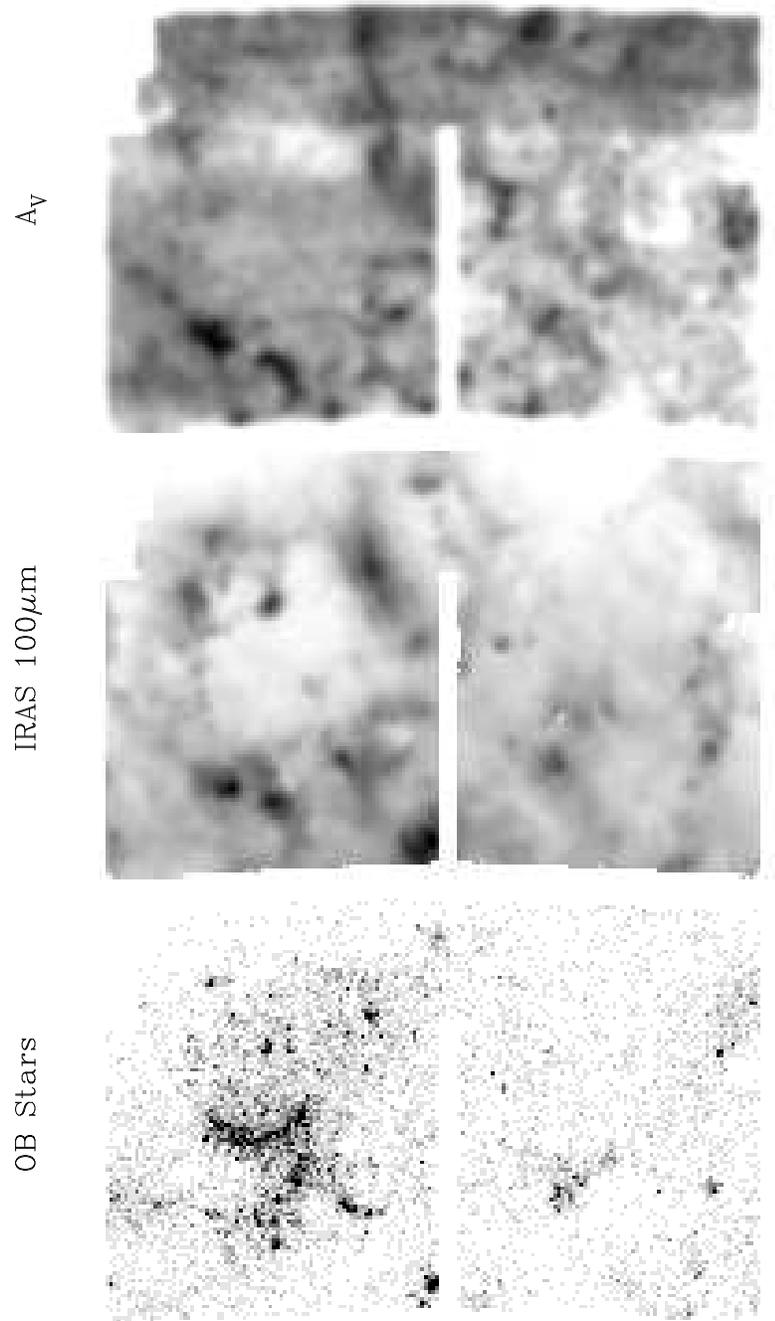}
\caption{
A comparison of the $A_V$ map derived from high $T_E$ stars (smoothed
to match the IRAS spatial resolution),
an IRAS 100$\mu$m image (log scaled), and a stellar density image of 
main sequence stars with $V<16.5$ and $B-V< 0.5$. 
The vertical bar at the right of the upper panel indicates the
scan that has non-photometric $I$-band data.}
\end{figure}
\clearpage

\begin{figure}
\vskip 5in
\includegraphics{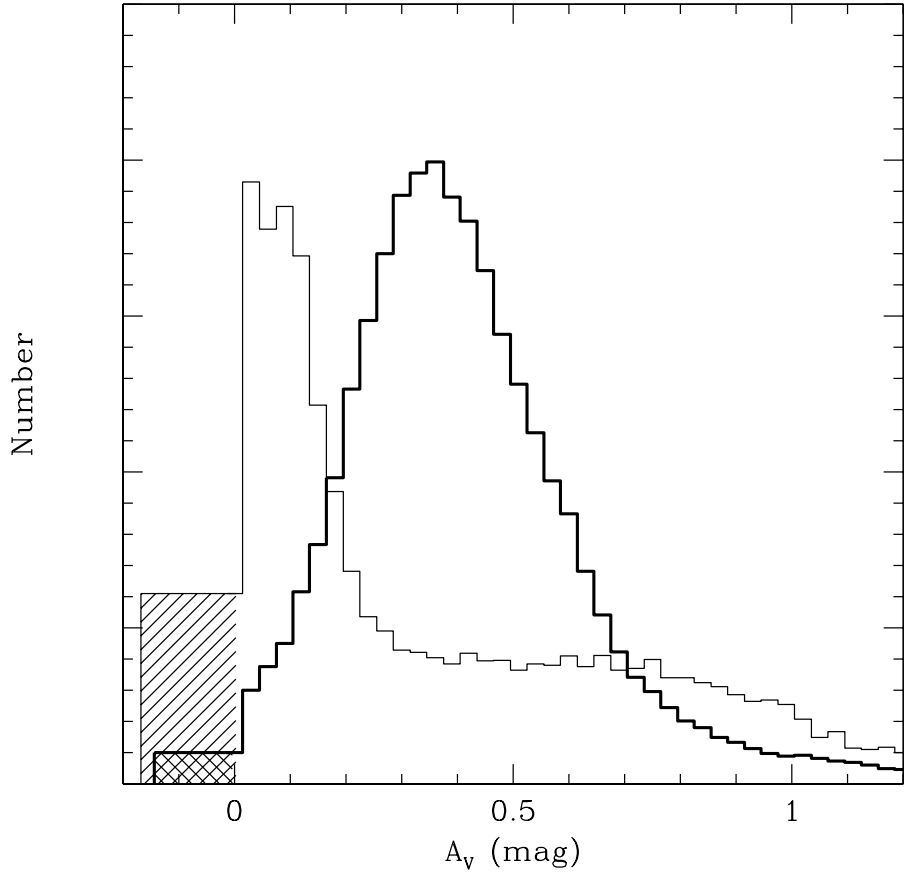}
\caption{
A comparison of the distribution of $A_V$'s derived for the
hot ($T_E > 12000$ K; thick solid line) and cool ($5500 < T_E < 6500$
K;
thin solid line) stars. The dashed areas at $A_V < 0$ indicate the 
number of stars with $A_V = 0$, where we have set the height of the
bin equal to the number of stars with $A_V = 0.01$ and the area 
to correctly represent the number of stars. The two histograms have
been
arbitrarily normalized.}
\end{figure}
\clearpage

\begin{figure}
\vskip 5in
\includegraphics{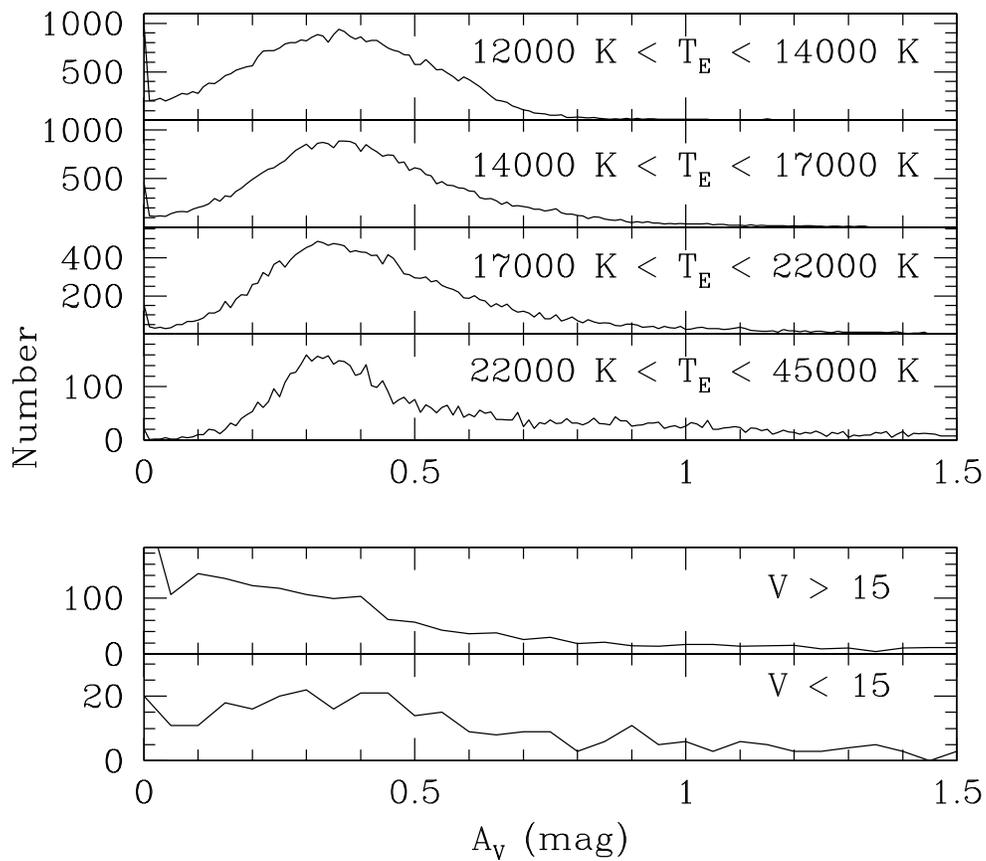}
\caption{
Comparison of extinction values for subsets of the high $T_E$
population. The lower 
two panels compare the values obtained by HZT using the
reddening-free method.}
\end{figure}
\clearpage

\begin{figure}
\vskip 5in
\includegraphics{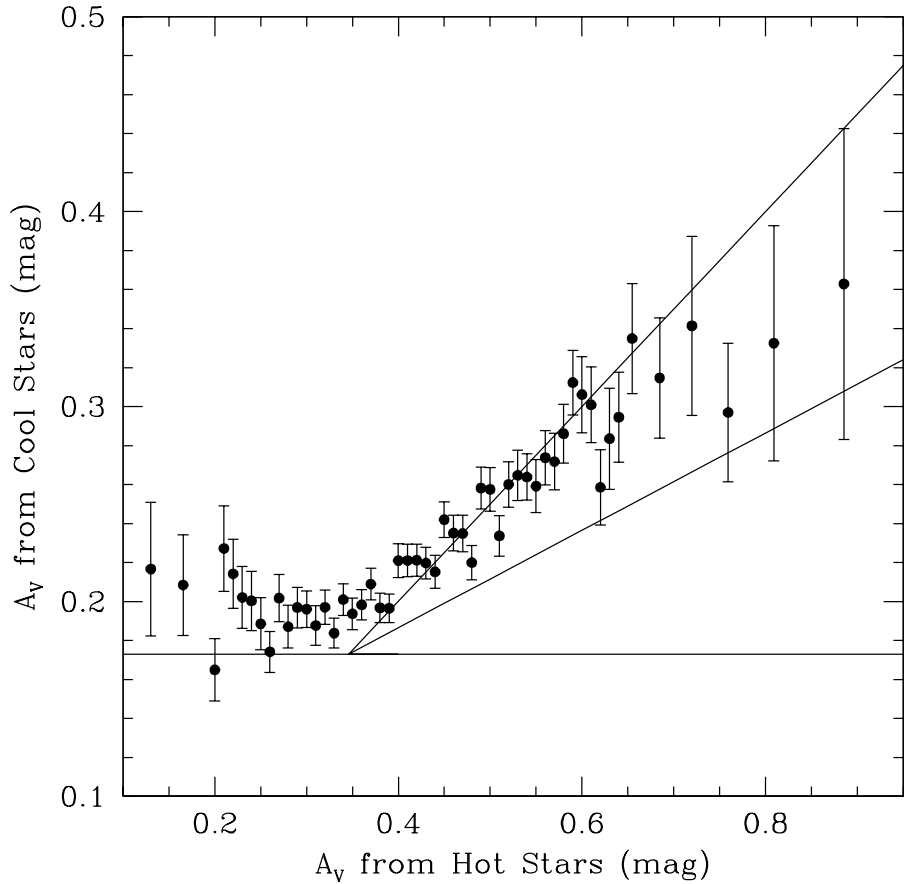}
\caption{
A comparison of the extinction derived for LMC stars with 
12000 K $< T_E <$ 45000 K
and stars with 5500 K $< T_E <$ 6500 K. The data are map pixel values
binned so that there are at least 200 pixels per bin. Errorbars represent 
standard deviation of the mean. The lines represent models as
described in text.}
\end{figure}
\clearpage

\end{document}